\documentclass[aps,showpacs,floatfix,notitlepage,superscriptaddress,longbibliography]{revtex4-1}
\usepackage{epsfig,graphics,amssymb,amsmath,subeqnarray,color,bm,bbm}
\usepackage[toc,page]{appendix}
\usepackage{mathtools}
\usepackage{xcolor}
\usepackage{float}
\usepackage{graphicx}
\usepackage[colorlinks=true,allcolors=blue]{hyperref}

\begin{document}
\title{Exact axisymmetric interaction of phoretically active Janus particles}
\author{Babak Nasouri}
\affiliation{Max Planck Institute for Dynamics and Self-Organization (MPIDS), 37077 Goettingen, Germany}
\author{Ramin Golestanian}
\email{ramin.golestanian@ds.mpg.de}
\affiliation{Max Planck Institute for Dynamics and Self-Organization (MPIDS), 37077 Goettingen, Germany}
\affiliation{Rudolf Peierls Centre for Theoretical Physics, University of Oxford, Oxford OX1 3PU, United Kingdom}
\date{\today}

\begin{abstract} 
{\color{black}{We study the axisymmetric interaction of two chemically active Janus particles. By relying on the linearity of the field equations and symmetry arguments, we derive a generic solution for the relative velocity of the particles. We show that regardless of the chemical properties of the system, the relative velocity can be written as a linear summation of geometrical functions which only depend on the gap size between the particles. We evaluate these functions via an exact approach which accounts for the full chemical and hydrodynamic interactions. Using the obtained solution, we expose the role of each compartment in the relative motion, and also discuss the contribution of different interactions. We then show that the dynamical system describing the relative motion of two Janus particles can have up to three fixed points. These fixed points can be stable or unstable, indicating that}} a system of two Janus particles can exhibit a variety of nontrivial behaviour depending on their initial gap size, and their chemical properties.
We also look at the specific case of Janus particles in which one compartment is inert, and present regime diagrams for their relative behaviour in the activity-mobility parameter space. 
\end{abstract}

\maketitle

\section{Introduction}
Phoretic transport has long been considered as a mechanism utilized by active particles for propulsion and navigation through an interactive medium \cite{gompper2020}. In this mechanism, which relies on nonequilibrium interfacial processes, the system exploits the inhomogeneity of its surrounding field and converts the free ambient energy into mechanical work \cite{anderson1989,anderson91}. This inhomogeneity can stem from a gradient in the chemical concentration \cite{derjaguin1947,RG-LesHouches}, temperature \cite{young1959,Golestanian:2012,cohen2014}, or electrostatic potential \cite{ramos1998,ajdari2000,bazant2004}, all of which can result in a net motion in the system.

Here, our focus is on diffusiophoretic processes, in which chemically active particles respond to a concentration gradient of chemicals, either imposed externally or induced by the particles themselves. The latter case, often referred to as self-diffusiophoresis, concerns a chemically active particle that can create a local perturbation in the concentration gradient via emitting or consuming chemicals through interfacial interactions {\color{black}{\cite{golestanian2005,hows07,simmchen2016,moran2017,RG-LesHouches,lohse2020}}}. 
{\color{black}{In the absence of advective effects which can lead to spontaneous symmetry breaking and directed motion in isotropic settings \cite{michelin2013,vajdi2020,chen2020}, an asymmetry in the concentration field is necessary to achieve autonomous motion or self-propulsion \cite{RG-LesHouches}.}} A well-known example of these self-propelling colloids are the Janus particles. These particles have (at least) two compartments with different physico-chemical properties, thereby inherently breaking the fore-aft symmetry \cite{golestanian2007}. The motion of a single Janus particle has been studied extensively, both theoretically and experimentally, and the underlying mechanism for its dynamical behaviour is well explored \cite{golestanian2005,ebbens2011,michelin2014,Ebbens2014,ibrahim2015,uspal2015,Ebbens2018}.

Pair interaction of phoretic particles has also been an immense topic of interest (see e.g., \citet{saha2014,saha2019} and \citet{sharifi2016} and the references therein). These interactions are of significant import in devising dimer-like micro- and nano-swimmers, wherein two phoretic particles are connected by a rod and propel autonomously by breaking the front-back symmetry \cite{reigh2015,ruckner2017,michelin2015b,michelin2017,reigh2018}. Furthermore, understanding these pair interactions can also be considered as the first step towards studying the suspension of phoretic particles, in which the system exhibits a variety of complex collective behaviours from swarming and comet-like propulsion, to phase separation and self-organization \cite{Liebchen2015,zottle2016,Colberg2017,stark2018,canalejo2019,varma2019}. Pair interactions also play a key role in resolving many-body interactions, since in these systems the near-field effects are often taken into account only through pair-wise interactions \cite{brady1988,varma2018}. Chemotaxis of enzymes can also be described via pair interactions, highlighting the importance of these interactions even at the molecular level \cite{illien2017b,canalejo2018,adeleke2019}.

Despite all of these, our understanding of the relative motion of two phoretic particles is still limited. One reason is that, due to complexity of the field equations, pair interactions are often modelled using far-field approximations \cite{varma2019,saha2019,burelbach2019}, which assumes the gap between the particles to be considerably larger than their length scale. Under this approach, the behaviour of the system cannot be probed when the particles are in close proximity of one another, and so the role of near-field chemical and hydrodynamic interactions cannot be explored. In an analytical/numerical approach, by using an exact approach, \citet{sharifi2016} looked at the pair interaction of two identical phoretic particles, taking into account the full chemical and hydrodynamic interactions. For chemically-identical Janus particles, they showed that the two particles can collapse, escape each other, or cease motion and become stationary. In this study, by allowing the particles to be of different chemical properties, we show that there are several more scenarios for the relative motion of two Janus particles. 
\color{black}
To solely focus on the chemical interplay between the compartments, we consider axisymmetric cases in which the particles can only translate along their common axis of symmetry. By using this simplification, and by extending the theoretical framework we developed for isotropic particles \cite{nasouri2020}, we derive a generic solution for the relative motion of two Janus particles of arbitrary chemical properties. This solution is in terms of compartment-by-compartment interactions which not only allows us to expose the contribution of each compartment to the relative behaviour, but also enables us to explore the full chemical parameter space. Both of these analyses are direct consequences of decomposing the chemical field, which is an approach that has not been employed before in phoretic systems.
\color{black}
Using the obtained generic solution, we show how the dynamical system describing the relative motion of two Janus particles can remarkably have up to three fixed points. Depending on the stability of these fixed points and the initial gap size between the particles, we discuss how the system can exhibit a wealth of nontrivial behaviours.

We begin by writing down the field equations governing the motion of two Janus particles with arbitrary chemical properties. We assume each Janus particle has two compartments (or faces) of equal coverage, and each compartment has its own chemical activity {\color{black}{and mobility. As the first step, we evaluate the interactions of the particles which have uniform mobilities. Using an exact analytical framework, we find a generic solution for the field equations, and analyze the relative motion in the full chemical parameter space, discussing the role of different compartments. We then use that solution to discuss the emergence of fixed points in the dynamical system representing the relative motion, and provide regime diagrams in the case of half-coated  (one compartment of each particle is completely inert) particles in the activity-mobility parameter space. We finally discuss the general case in which the mobilities of the two compartments can also differ, and present a generic solution for that case as well.}}

\section{Problem Statement}
\begin{figure}
\begin{center}
\includegraphics[width=0.9\columnwidth]{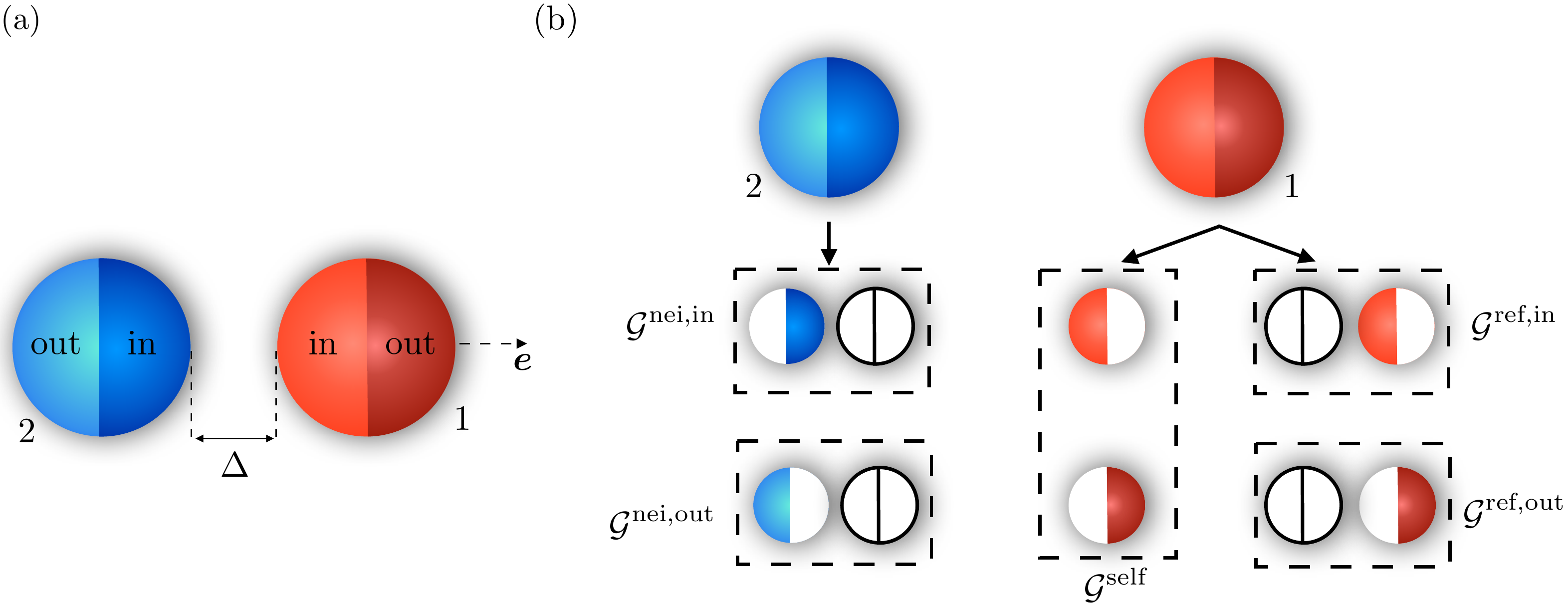}
\caption{(a) Schematic of the two Janus particles considered in this study. Each particle has two equally-sized compartments. We label the compartments facing each other using `in', and use `out' to describe the outer ones. The unit vector $\bm{e}$ is the common axis of symmetry, and $\Delta$ is the clearance between the particles. {\color{black}{(b) Schematic of the chemical field decomposition to isolated self-propulsion ($\mathcal{G}^\text{self}$), neighbour-induced interaction ($\mathcal{G}^\text{nei,in},\mathcal{G}^\text{nei,out}$), and neighbour-reflected ones ($\mathcal{G}^\text{ref,in},\mathcal{G}^\text{ref,out}$) from the perspective of particle 1.}} 
}
\label{scheme}
\end{center}
\end{figure}

We consider two spheres (sphere 1 and sphere 2) of radii $R$ and gap size $\Delta$, immersed in an otherwise quiescent viscous fluid. The system is axisymmetric, and we define a unit vector $\bm{e}$ as the axis of symmetry. These spheres are chemically active, and they interact with a chemical (i.e., solute particles) of diffusion coefficient $D$. In the infinite dilution limit of solute particles, and in the absence of any nearby boundaries or a background concentration gradient, the relative concentration field can be expressed by a steady-state diffusion equation
\begin{align}
\nabla^2 C=0.
\end{align}
Here, we have assumed the advective effects in the solute transport to be negligible compared to the diffusive effects (i.e. P\`eclet number is vanishingly small). The spheres perturb the concentration field by consuming/producing the solute particles, thereby creating a normal flux at their surfaces (i.e., $\mathcal{S}_1$ and $\mathcal{S}_2$). We may write
\begin{align}
D\bm{n}_1\cdot\bm{\nabla}C&=-\alpha_1\quad\text{at}~\mathcal{S}_1,\quad\quad D\bm{n}_2\cdot\bm{\nabla}C=-\alpha_2\quad\text{at}~\mathcal{S}_2,
\end{align}
where $\bm{n}_1$ and $\bm{n}_2$ are unit vectors normal to the surfaces, and $\alpha_1$ and $\alpha_2$ are the catalytic activities of sphere 1 and sphere 2, respectively. The spheres respond to a gradient in the chemical field through interfacial interactions, characterized by a physico-chemical property called mobility. This response is often modelled as a local fluid slip velocity at the surface of each sphere, and can be written as
\begin{align}
\label{bc-chem}
\bm{v}_1^\text{s}&=\mu_1(\bm{I}-\bm{n}_1\bm{n}_1)\cdot\bm{\nabla}C\quad\text{at}~\mathcal{S}_1,\quad\quad\bm{v}_2^\text{s}=\mu_2(\bm{I}-\bm{n}_2\bm{n}_2)\cdot\bm{\nabla}C\quad\text{at}~\mathcal{S}_2,
\end{align}
where $\mu_1$ and $\mu_2$ are the mobilities of the particles. Here, we consider the {\color{black}{axisymmetric}} interactions of two Janus particles, as shown in figure~\ref{scheme}(a). \color{black}These particles have two equally-sized compartments with different coatings which may result in a discontinuity in their surface activity and mobility. We use `in' to describe the chemical properties of the compartments facing each other ($\alpha_1^\text{in}, \mu_1^\text{in},\alpha_2^\text{in},\mu_2^\text{in}$), and `out' for the outer compartments ($\alpha_1^\text{out}, \mu_1^\text{out},\alpha_2^\text{out},\mu_2^\text{out}$).\color{black}
The chemically induced slip velocities may give rise to translational motion of the spheres. In the absence of inertia (zero Reynolds number regime), one may find the velocities of each sphere, $\bm{V}_1$ and $\bm{V}_2$, by solving the Stokes equations
\begin{align}
\label{stokes1}
\eta\nabla^2\bm{v}&= \bm{\nabla}p,\quad\quad\bm{\nabla}\cdot\bm{v}=0,
\end{align}
subject to boundary conditions
\begin{align}
\label{stokes3}
\bm{v}\left(\bm{x}\in\mathcal{S}_1\right)&=\bm{V}_1+\bm{v}_1^\text{s},\quad\quad\bm{v}\left(\bm{x}\in\mathcal{S}_2\right)=\bm{V}_2+\bm{v}_2^\text{s},\quad\quad \bm{v}(|\bm{x}-\bm{x}_i|\rightarrow\infty)=\bm{0},
\end{align}
where $\bm{v}$ and $p$ are the velocity and pressure field, $\eta$ is the fluid viscosity, $\bm{x}$ is the position vector, and $\bm{x}_i$ denotes the centre of sphere $i$ with $i\in\{1,2\}$. Since the system is axisymmetric, the particles cannot rotate and only translate along the axis of symmetry.
\section{Non-uniform activity and uniform mobility}
\subsection{A generic solution}
\color{black}
To find the translational velocities of the spheres, we need to solve the chemical and hydrodynamic interactions which are coupled through the boundary conditions given in equations \eqref{bc-chem} and \eqref{stokes3}. However, we can simplify the calculations by using the symmetry arguments and relying on the linearity of the field equations, as we show in the following. 
\color{black}
To illustrate this approach, we first begin with the hydrodynamic interactions, for which we can use the Lorentz reciprocal theorem to bypass solving the complete Stokes equations \cite{Lorentz1896,happel1983,stone1996,elfring2017,nasouri2018,nasouri2018b,masoud2019}. This theorem connects our main problem, to an auxiliary one in the same domain as 
\begin{align}
\left<\bm{n}\cdot\bm{\sigma}\cdot\hat{\bm{v}}\right>_{\mathcal{S}_1+\mathcal{S}_2}=\left<\bm{n}\cdot\hat{\bm{\sigma}}\cdot{\bm{v}}\right>_{\mathcal{S}_1+\mathcal{S}_2},
\end{align}
where $\left<\cdot\right>$ denotes the surface integral, and $\bm{n}$ is a unit vector normal to the surface of the domain. Here ($\bm{\sigma},\bm{v}$) and ($\hat{\bm{\sigma}},\hat{\bm{v}}$) are the stress and velocity fields in the main and auxiliary problem, respectively. By choosing the auxiliary problem as the axisymmetric motion of two passive particles (with the same geometry as in our main problem) towards each other with an identical and constant speed, we can directly find the relative velocity in terms of the flow properties of the auxiliary problem \cite{sharifi2016,papavassiliou2017,yang2019}. Defining $\hat{\bm{F}}_i$ as the net hydrodynamic force on each particle in the auxiliary problem, the relative velocity in the main problem is then found
\begin{align}
\label{relative}
\bm{V}_1-\bm{V}_2=\frac{\bm{e}}{|\hat{\bm{F}}_1|}\left(\left<\hat{\sigma}_{1}{v}_1^{\rm s}\right>_{{\cal S}_1}+\left<\hat{\sigma}_{2}{v}_2^{\rm s} \right>_{{\cal S}_2}\right),
\end{align}
where $\hat{\sigma}_{i}=\bm{n}_i\cdot\hat{\bm{\sigma}}\cdot\bm{t}_i$ is the tangential component of the normal traction, $\bm{v}^{\rm s}_i=v^{\rm s}_i\bm{t}_i$, and $\bm{t}_i$ is a unit vector tangential to the surface of sphere~$i$.
\color{black}
We note that since we are only interested in the relative motion of the particles, and that the particles are of equal radii, it suffices to employ only one auxiliary problem to resolve the hydrodynamic interactions. For instance, probing the velocities of the individual particles requires an additional auxiliary problem, which is often chosen to be the trailing of two passive particles in a viscous fluid \cite{stimson1926}. When the system is not axisymmetric and the particles freely move with respect to one another, even further decomposition of the hydrodynamic field is required as one needs to account for parallel and perpendicular translational motions, as well as rotations. A detailed analysis with respect to this sort of decomposition in the hydrodynamic field is presented by \citet{mozaffari2016} and \citet{sharifi2016}. Here, however, we focus instead on decomposing the chemical field as we want to better understand the contribution of each compartment separately in the relative behaviour of the particles. To this end, we limit our attention to the relative motion in an axisymmetric setting, so that the hydrodynamic interactions can be probed by simply using one single auxiliary problem. As we will show in the following, this simplification allows us to decompose the overall interactions to some geometrical functions that can fully capture the dynamics of the system.

\color{black}
We now use equation \eqref{relative} to decompose the interactions in the chemical field. Without any loss of accuracy, the concentration field can be written as
\begin{align}
C(\bm{x})=C_1(\bm{x})+C_2(\bm{x}),
\end{align}
where $C_1$ ($C_2$) is the concentration field induced by sphere 1 (2) when sphere 2 (1) is completely inert. The concentration field can be further decomposed as
\begin{align}
C(\bm{x})=\left[C^\text{far}_1(\bm{x})+C^\text{near}_1(\bm{x})\right]+\left[C^\text{far}_2(\bm{x})+C^\text{near}_2(\bm{x})\right],
\end{align}
where `far' denotes the concentration field induced by each particle in the absence of its neighbour, and `near' accounts for the correction due to the chemical interactions between the particles. The slip velocity for each sphere is then found
 \begin{align}
\label{slip}
\bm{v}_i^\text{s}=\mu_i\left[\bm{\nabla}^i_\parallel C^\text{far}_1+\bm{\nabla}^i_\parallel C^\text{near}_1\right]+\mu_i\left[\bm{\nabla}^i_\parallel C^\text{far}_2+\bm{\nabla}^i_\parallel C^\text{near}_2\right]\quad \text{at}~\mathcal{S}_i,
\end{align}
where $\bm{\nabla}^i_\parallel=\left(\bm{I}-\bm{n}_i\bm{n}_i\right)\cdot\bm{\nabla}$.
\color{black}
Note that we have not yet used the assumption of $\mu_i^\text{in}=\mu_i^\text{out}$, so the decomposition given in equation \eqref{slip} is generic. But to simplify the equations even further, we now assume that the mobilities of the spheres do not vary across their surfaces. By replacing the slip velocity from equation \eqref{slip} to \eqref{relative}, we can make some simplifications. 
\color{black}The motion induced by $C_i^\text{far}$ is essentially self-propulsion in the absence of any neighbours. Thus it must linearly depend on $\alpha_i^\text{in}-\alpha_i^\text{out}$, so one can claim 
\begin{align}
\frac{D}{|\hat{\bm{F}}_1|}\left<\hat{\sigma}_i\bm{\nabla}^i_\parallel C^\text{far}_i\right>_{\mathcal{S}_i}&=\bm{e}\mathcal{G}^\text{self}_i\left(\alpha_i^\text{in}-\alpha_i^\text{out}\right),
\end{align}
where $\mathcal{G}^\text{self}_i$ only varies with the cap size \cite{golestanian2007}. Note that when a particle is chemically isotropic ($\alpha_i^\text{in}=\alpha_i^\text{out}$), it cannot self-propel without the presence of a nearby neighbouring particle or boundary since its concentration field becomes completely isotropic \cite{soto2014}. We can similarly define
\begin{align}
\frac{D}{|\hat{\bm{F}}_1|}\left<\hat{\sigma}_i\bm{\nabla}^i_\parallel C^\text{near}_i\right>_{\mathcal{S}_i}&=\bm{e}\mathcal{G}^\text{ref,in}_i\alpha_i^\text{in}+\bm{e}\mathcal{G}^\text{ref,out}_i\alpha_i^\text{out},\\
\frac{D}{|\hat{\bm{F}}_1|}\left<\hat{\sigma}_i\bm{\nabla}^i_\parallel C_j\right>_{\mathcal{S}_i}&=\bm{e}\mathcal{G}^\text{nei,in}_i\alpha_j^\text{in}+\bm{e}\mathcal{G}^\text{nei,out}_i\alpha_j^\text{out},
\end{align}
where all the `$\mathcal{G}$' functions are dimensionless and only depend on the gap size, and $\{i,j\}\in\{1,2\}$ in a mutually-exclusive manner. Here, {\color{black}{as shown schematically in figure~\ref{scheme}(b)}}, $\mathcal{G}^\text{ref,in}_i$ and $\mathcal{G}^\text{ref,out}_i$ represent the motion induced by the chemical activity of a particle, due to the passive presence of its neighbour. Thus, in these terms, the neighbouring particle serves as a geometrical asymmetry in the concentration field generated by each particle. Note that since the two compartments of each particle interact differently with the neighbouring particle, $\mathcal{G}^\text{ref,in}_i\neq\mathcal{G}^\text{ref,out}_i$ specially when the gap size is small. On the other hand, $\mathcal{G}^\text{nei,in}_i$ and $\mathcal{G}^\text{nei,out}_i$ account for the motions induced solely by the chemical field of the neighbouring particle. Similarly here, $\mathcal{G}^\text{nei,in}_i\neq\mathcal{G}^\text{nei,out}_i$.
Due to the symmetry of the system, for all the $\mathcal{G}$ functions we find $\left(\mathcal{G}_i\right)^*=\mathcal{G}_j\equiv\mathcal{G}$, where $(\cdot)^*$ denotes a mirror-symmetric transformation. Defining the relative speed as $V_\text{rel}=\left(\bm{V}_1-\bm{V}_2\right)\cdot{\bm{e}}$, we finally arrive at
\begin{align}
\label{relative2}
V_\text{rel}=&\mathcal{G}^\text{self}\left[\mu_1\left(\alpha_1^\text{in}-\alpha_1^\text{out}\right)+\mu_2\left(\alpha_2^\text{in}-\alpha_2^\text{out}\right)\right]/D+\nonumber\\
&\mathcal{G}^\text{nei,in}\left(\mu_1\alpha_2^\text{in}+\mu_2\alpha_1^\text{in}\right)/D+\mathcal{G}^\text{ref,in}\left(\mu_1\alpha_1^\text{in}+\mu_2\alpha_2^\text{in}\right)/D+\nonumber\\
&\mathcal{G}^\text{nei,out}\left(\mu_1\alpha_2^\text{out}+\mu_2\alpha_1^\text{out}\right)/D+\mathcal{G}^\text{ref,out}\left(\mu_1\alpha_1^\text{out}+\mu_2\alpha_2^\text{out}\right)/D.
\end{align}
Equation $\eqref{relative2}$ presents a generic expression for the relative speed for any two Janus particles. It shows that the relative motion of the particles is governed by their self-propulsion ($\mathcal{G}^\text{self}$), neighbour-induced motions ($\mathcal{G}^\text{nei,in}$ and $\mathcal{G}^\text{nei,out}$), and self-generated neighbour-reflected motions ($\mathcal{G}^\text{ref,in}$ and $\mathcal{G}^\text{ref,out}$). The geometrical $\mathcal{G}$ functions are independent of the chemical properties of the particles, thus we only need to evaluate them once. Contrary to equation \eqref{relative} wherein the chemical and hydrodynamic fields are both needed to be solved upon variation of the chemical properties, equation \eqref{relative2} allows us to determine the relative velocity quite efficiently using just a simple linear summation of particles' chemical properties and the geometrical functions.
\color{black}
{\color{black}{Except for the case of self-propulsion \cite{golestanian2007}, finding the exact explicit analytical expressions for each of these geometrical functions may not be feasible. However, one can evaluate them to some level of approximation using the far-field based approaches such as the method of reflections, or employ the exact treatment which presents the solution in terms of infinite series in the bispherical coordinates \cite{mozaffari2016,sharifi2016,michelin2017}. Here, to be able to see the relative behaviour without any loss of accuracy, we take the latter approach and numerically evaluate these functions accounting for the full chemical and hydrodynamic interactions.}} This approach, which relies on the reciprocal theorem and the exact solution of the Laplace and Stokes equation for a two-body system, has been employed in the literature quite extensively to evaluate the motion of phoretic particles \cite{mozaffari2016,sharifi2016,michelin2017,nasouri2020}. In what follows we show how this method can be adapted to obtain the geometrical functions. 
 
\begin{figure}
\begin{center}
\includegraphics[width=\columnwidth]{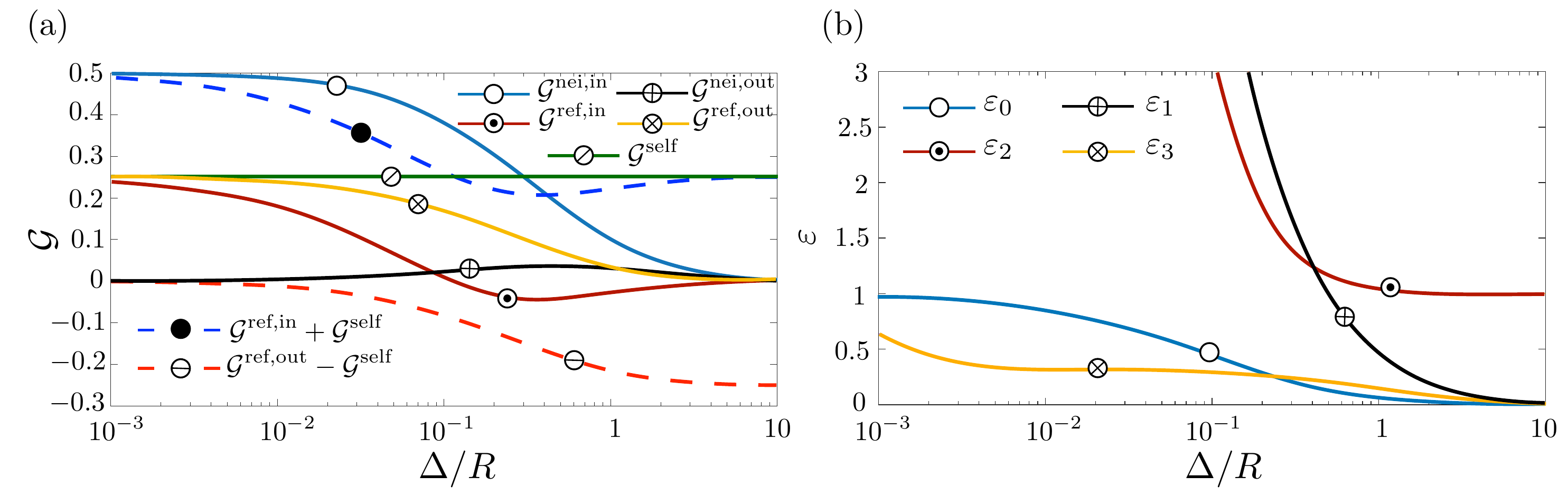}
\caption{\color{black}(a) Variation of the geometrical $\mathcal{G}$ functions against the gap size. As shown in equation \eqref{relative2}, the relative velocity of the particles can be expressed as a linear summation of these functions. (b) The ratio of the re-grouped $\mathcal{G}$ functions which decay monotonically with $\Delta$ as defined in equations \eqref{iso} to \eqref{eps_last}.}
\label{gs}
\end{center}
\end{figure}

\begin{figure}
\begin{center}
\includegraphics[width=0.5\columnwidth]{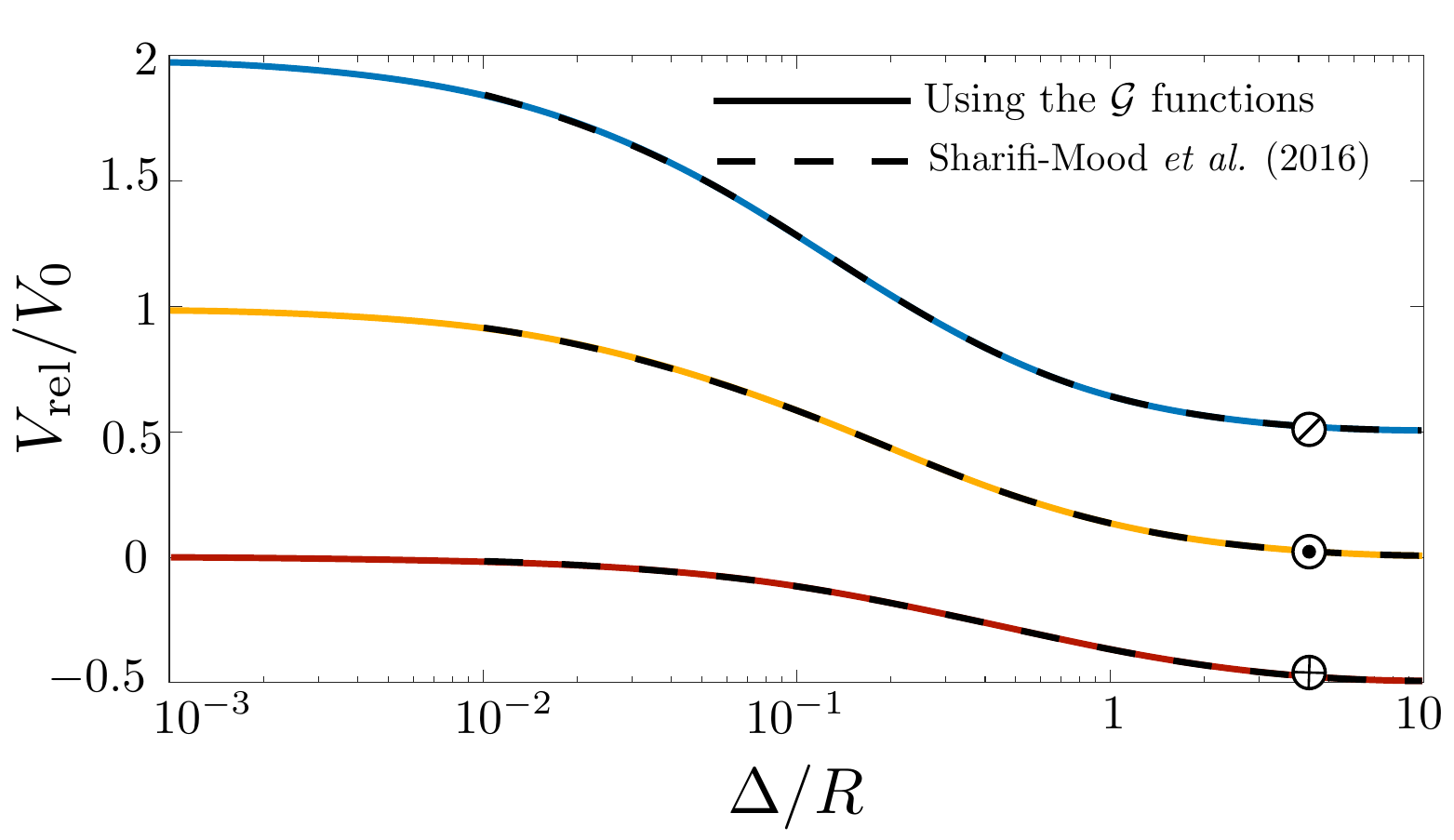}
\caption{\color{black}{The relative speed obtained from equation \eqref{relative2} (solid lines) and those obtained by \citet{sharifi2016} (dashed-lines) for $10^{-3}<\Delta/R<10$. In all cases $\tilde{\mu}_1=\tilde{\mu}_2=1$, and $\oslash$: $\tilde{\alpha}_1^\text{in}=\tilde{\alpha}_2^\text{in}=1$, $\tilde{\alpha}_1^\text{out}=\tilde{\alpha}_2^\text{out}=0$, $\odot$: $\tilde{\alpha}_1^\text{in}=\tilde{\alpha}_2^\text{out}=1$, $\tilde{\alpha}_1^\text{out}=\tilde{\alpha}_2^\text{in}=0$, and $\oplus$: $\tilde{\alpha}_1^\text{in}=\tilde{\alpha}_2^\text{in}=0$, $\tilde{\alpha}_1^\text{out}=\tilde{\alpha}_2^\text{out}=1$.}}
\label{compare}
\end{center}
\end{figure}

For the case of uniform mobilities, we need to evaluate five geometrical functions. Since the system is linear, if we find the relative speed for five arbitrarily-chosen cases (i.e., five pair interactions with arbitrarily chosen values for activities and mobilities), we can construct a linear system of equations from which the exact values for the $\mathcal{G}$ functions can be recovered. To do so, we use equation \eqref{relative} which describes the relative speed of the particles in terms of the slip velocities and the flow field of the auxiliary problem. For any pair of spherical particles, we can solve the chemical field equations exactly in the bispherical coordinate system, as widely discussed in the literature \cite{michelin2015b,mozaffari2016,sharifi2016}. The complete solution to the auxiliary problem is also readily available from the classical works of \citet{maude1961} and \citet{spielman1970}. Thus, combining these two, the exact relative velocity of the particles can be explicitly determined from equation \eqref{relative}. Using this direct approach, we can construct a $5\times5$ matrix which can be used to determine the $\mathcal{G}$ functions.
We take $\alpha_0$ and $\mu_0$ as the reference values for the activity and mobility, and define $\tilde{\alpha}=\alpha/\alpha_0$, and $\tilde{\mu}=\mu/\mu_0$. The scaling for the speed then naturally arises as $V_0=\alpha_0\mu_0/D$, which essentially characterizes the self-propulsion speed of a half-coated Janus particle with chemical properties $\alpha_0$ and $\mu_0$, as such a particle would swim with the speed of $(1/4)V_0$ \cite{golestanian2007}. Using these scalings, we evaluate the geometrical functions for $0.001<\Delta/R<10$; see figure~\ref{gs}(a). {\color{black}{ To validate the obtained values, we use them to determine the relative speed for Janus particles with identical chemical properties, and as shown in figure~\ref{compare}, our results concisely match those reported by \citet{sharifi2016}}}.

\color{black}As expected, the function $\mathcal{G}^\text{self}_i$ which represents the isolated self-propulsion, does not vary with the gap size and is solely a function of the coating ratio between the two compartments (which we consider here to be $1$ as each compartment takes a half of the surface). We find $\mathcal{G}^\text{self}_i=0.25$ which is identical to the exact value obtained analytically for a single Janus particle \cite{golestanian2007}. The other $\mathcal{G}$ functions, however, originate from the chemical and hydrodynamic interactions between the particles. In the far-field limit, the chemical field generated by each particle can be approximated by a point source, while the hydrodynamic field is the one of a force-free torque-free motion which is governed by a symmetric force dipole (i.e. Stresslet). Thus, in this limit, the chemical field generated by each particle will decay by $1/\Delta$, while the hydrodynamic field will decay as $1/\Delta^2$. Therefore, the net phoretic interactions of the two particles will also decay by the gap size, as can be seen from the behaviour of the geometrical functions when $\Delta$ is large. Remarkably, however, this weakening of interactions does not occur monotonically for $\mathcal{G}^\text{nei,out}$ and $\mathcal{G}^\text{ref,in}$. For the former, an increase in $\Delta$ initially strengthens the interactions, while for the latter the attractive/repulsive nature of the interactions is reversed at a certain gap size. This implies that the near-field chemical and hydrodynamic interactions in these geometrical functions may oppose the leading-order far-field effects. When the gap size is small, the strength of the near-field effects dominates the interactions and results in the non-monotonicity of the interactions with respect to the gap size. These effects rapidly vanish once $\Delta/R>1$, and the interactions follow a monotonic decay as dictated by the far-field interactions. 

We recall that the expression given in equation \eqref{relative2} is derived using the linearity of the field equations and the geometrical symmetry arguments. {\color{black}{Thus, while the full behaviour can only be explored via an exact approach such as the one taken in this study, one can still find these geometrical functions to some level of approximations using the far-field based approaches such as the method of reflections.}} The method of reflections assumes the gap size to be considerably larger than the length scale of the particles. At the zeroth order, the chemical field generated by each particle is substituted by a point source (or sink depending on the sign of activity), while the hydrodynamic interactions are completely ignored. At this order, the neighbour-reflected terms are identically zero, as they only appear at higher orders. Further reflections then account for higher-order effects such as source-doublets in the chemical field, and stresslets in the hydrodynamic field, giving rise to the appearance of the neighobour-reflected terms and also correcting the neighbour-induced ones. By keeping more reflections, the solution eventually converges to that of the exact approach, but only in the limit of $\Delta/R>1$ {\color{black}{(see e.g. \citet{sharifi2016} for the comparison)}}. Thus, we can technically recover the geometrical functions using the method of reflections with reasonable accuracy for $\Delta/R>1$. However, when $\Delta\sim R$, the convergence becomes very slow and one has to account for several reflections.

It is now worthwhile to discuss the case wherein the particles are in very close proximity of one another and $\Delta\rightarrow0$. One then naturally expects the effect of the outer compartments to become vanishingly small compared to those of the inner ones. To evaluate this, we need to separate the effects of the inner and outer compartments completely. We note that the decomposition given in equation \eqref{relative2} does not properly separate the role of each compartment; rather it shows how each type of interaction contributes to the relative motion. Thus, to evaluate the role of each compartment, we need to decompose the self-propulsion term as well. We thereby combine $\mathcal{G}^\text{self}$ with the neighbour-reflected terms, namely $\mathcal{G}^\text{ref,in}$ and $\mathcal{G}^\text{ref,out}$. The total number of the geometrical functions then reduces to four as we can write
\begin{align}
\label{combined}
V_\text{rel}=&\mathcal{G}^\text{nei,in}\left(\mu_1\alpha_2^\text{in}+\mu_2\alpha_1^\text{in}\right)/D+\left(\mathcal{G}^\text{ref,in}+\mathcal{G}^\text{self}\right)\left(\mu_1\alpha_1^\text{in}+\mu_2\alpha_2^\text{in}\right)/D+\nonumber\\
&\mathcal{G}^\text{nei,out}\left(\mu_1\alpha_2^\text{out}+\mu_2\alpha_1^\text{out}\right)/D+\left(\mathcal{G}^\text{ref,out}-\mathcal{G}^\text{self}\right)\left(\mu_1\alpha_1^\text{out}+\mu_2\alpha_2^\text{out}\right)/D.
\end{align}
As shown in figure~\ref{gs}(a), the total contribution of the outer compartments (i.e., $\mathcal{G}^\text{nei,out}$ and $\mathcal{G}^\text{ref,out}-\mathcal{G}^\text{self}$) indeed asymptotes to zero when $\Delta\rightarrow0$, while those of the inner compartments reach finite values. One can then conclude that in the lubrication regime, both the chemical and hydrodynamic interactions of the outer compartments are fully screened over the spherical boundary of the particles. A similar screening was observed for hydrodynamic interactions of two beating cilia, for which the presence of a spherical boundary was shown to fully screen the interactions \cite{nasouri2016}. Finally, we note that when $\Delta\rightarrow0$, we find $\mathcal{G}^\text{nei,in}$ to be two times larger than $\mathcal{G}^\text{self}$. This suggests that in the lubrication regime, the squeezing effect on the chemical and hydrodynamic field strengthens the overall phoretic interactions such that the neighbouring particle can remarkably translate the particle faster than its own inherent asymmetry. 
\color{black}
\subsection{Emergence of fixed points}
\color{black}
Depending on the chemical properties of the particles (and also $\Delta$ in the case of $\mathcal{G}^\text{ref,in}$) the interactions stemmed from the geometrical functions can be attractive or repulsive. Therefore, they may (or may not) oppose one another in a given pair interaction. Additionally, since these geometrical functions (except for the one of the self-propulsion) vary with the gap size, the overall nature of the phoretic interaction may also vary with the gap size. Thus, the collective interplay of all these effects may induce fixed points in the dynamical behaviour of the system.
\begin{figure}
\begin{center}
\includegraphics[width=\columnwidth]{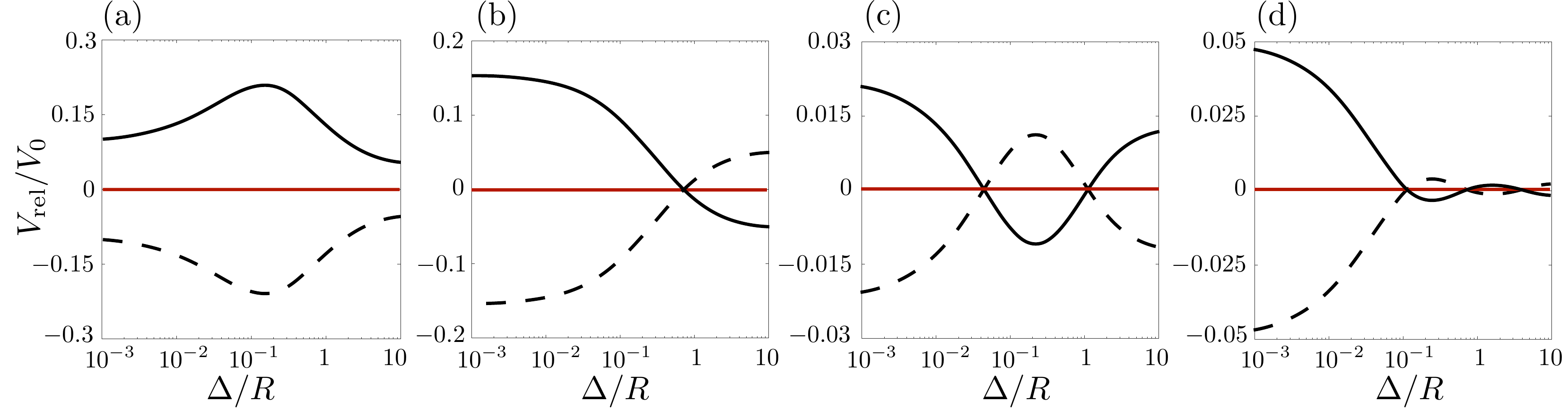}
\caption{Variation of the relative speed ($V_\text{rel}$) with the gap size for four different cases. The dynamical system describing the relative motion of the two Janus particles can have (a) zero, (b) one, (c) two, or (d) three fixed points. The parameter sets used for solid lines are as follows: (a) $\tilde{\alpha}_1^\text{in}=-0.82$, $\tilde{\alpha}_1^\text{out}=-0.84$, $\tilde{\alpha}_2^\text{in}=0.56$, $\tilde{\alpha}_2^\text{out}=0.81$, $\tilde{\mu}_1=0.07$, $\tilde{\mu}_2=-0.78$, (b) $\tilde{\alpha}_1^\text{in}=-0.8$, $\tilde{\alpha}_1^\text{out}=-0.64$, $\tilde{\alpha}_2^\text{in}=-0.28$, $\tilde{\alpha}_2^\text{out}=-0.89$, $\tilde{\mu}_1=0.04$, $\tilde{\mu}_2=-0.33$, (c) $\tilde{\alpha}_1^\text{in}=-0.26$, $\tilde{\alpha}_1^\text{out}=-0.47$, $\tilde{\alpha}_2^\text{in}=0.37$, $\tilde{\alpha}_2^\text{out}=0.26$, $\tilde{\mu}_1=0.05$, $\tilde{\mu}_2=0.37$, and (d) $\tilde{\alpha}_1^\text{in}=0.89$, $\tilde{\alpha}_1^\text{out}=-0.16$, $\tilde{\alpha}_2^\text{in}=-0.79$, $\tilde{\alpha}_2^\text{out}=0.90$, $\tilde{\mu}_1=0.58$, $\tilde{\mu}_2=0.37$. The same values are used for the dashed lines except $\mu_1\rightarrow -\mu_1$ and $\mu_2\rightarrow -\mu_2$. The red solid lines show the value zero.}
\label{v_rel_plot}
\end{center}
\end{figure}

To explore this, we first look at the simple case of chemically isotropic particles, for which it was shown that the relative motion can only have one fixed point \cite{nasouri2020}. Note that in this case there is no self-propulsion as the chemical field generated by each particle in isolation is purely isotropic. Also, since there is no difference between the two compartments of each particle, we can group the geometrical functions together and define $\mathcal{G}^\text{nei}=\mathcal{G}^\text{nei,in}+\mathcal{G}^\text{nei,out}$ as the net neighbour-induced interaction, and $\mathcal{G}^\text{ref}=\mathcal{G}^\text{ref,in}+\mathcal{G}^\text{ref,out}$ as the net neighobour-reflected contribution. By setting $\alpha_i^\text{in}=\alpha_i^\text{out}$, the relative speed given in equation \eqref{relative2} takes the simple form
\begin{align}
\label{iso}
V_\text{rel}=\mathcal{G}^\text{nei}\left[\left(\mu_1\alpha_2+\mu_2\alpha_1\right)+\varepsilon_{0}\left(\mu_1\alpha_1+\mu_2\alpha_2\right)\right]/D,
\end{align}
where 
\begin{align}
\varepsilon_{0}=\frac{\mathcal{G}^\text{ref}}{\mathcal{G}^\text{nei}}. 
\end{align}
As we discussed in our previous work \cite{nasouri2020}, $\mathcal{G}^\text{nei}$ and $\mathcal{G}^\text{ref}$ are both positive scalers that decay monotonically with the gap size, but so does their ratio; see figure~\ref{gs}(b). Thus, the system of two isotropic particles can indeed have at most one fixed point.

We now want to similarly determine the number of fixed points in a pair interaction of Janus particles. Since the interactions are more complex for Janus particles, a simple regrouping of the geometrical function may not suffice for unraveling the dynamical system. However, with the aid of numerical calculations, we can show that the system can only have three fixed points. We rewrite equation \eqref{relative2} as
\begin{align}
\label{relative3}
V_\text{rel}=\left(\mathcal{G}^\text{self}-\mathcal{G}^\text{ref,out}\right)\bigg[&\left(\mu_1\alpha_2^\text{in}+\mu_2\alpha_1^\text{in}\right)\varepsilon_1+\left(\mu_1\alpha_1^\text{in}+\mu_2\alpha_2^\text{in}\right)\varepsilon_2+\left(\mu_1\alpha_2^\text{out}+\mu_2\alpha_1^\text{out}\right)\varepsilon_3-\left(\mu_1\alpha_1^\text{out}+\mu_2\alpha_2^\text{out}\right)\bigg]/D,
\end{align}
where 
\begin{align}
\varepsilon_1&=\frac{\mathcal{G}^\text{nei,in}}{\mathcal{G}^\text{self}-\mathcal{G}^\text{ref,out}},\\
\varepsilon_2&=\frac{\mathcal{G}^\text{ref,in}+\mathcal{G}^\text{self}}{\mathcal{G}^\text{self}-\mathcal{G}^\text{ref,out}},\\
\varepsilon_3&=\frac{\mathcal{G}^\text{nei,out}}{\mathcal{G}^\text{self}-\mathcal{G}^\text{ref,out}},
\label{eps_last}
\end{align}
 are now all positive scalars that decay monotonically with $\Delta$, as shown in figure~\ref{gs}(b). Given that $\mathcal{G}^\text{self}-\mathcal{G}^\text{ref,out}$ is always positive, the nature of the interactions is now only embedded in the pre-factors containing the chemical properties of the particles, and so determining the number of the fixed point is reduced to the terms inside the bracket in equation \eqref{relative3}. A simple parameter scan then reveals that the dynamical system allows for a maximum of three fixed points. Unlike the case of isotropic particles, the emergence of the fixed points here is not solely due to a simple interplay of neighbour-induced and neighbour-reflected interactions. Rather, as shown in equation \eqref{relative3}, a combination of different interactions lead to emergence of the fixed points.

\color{black}
\begin{figure}
\begin{center}
\includegraphics[width=\columnwidth]{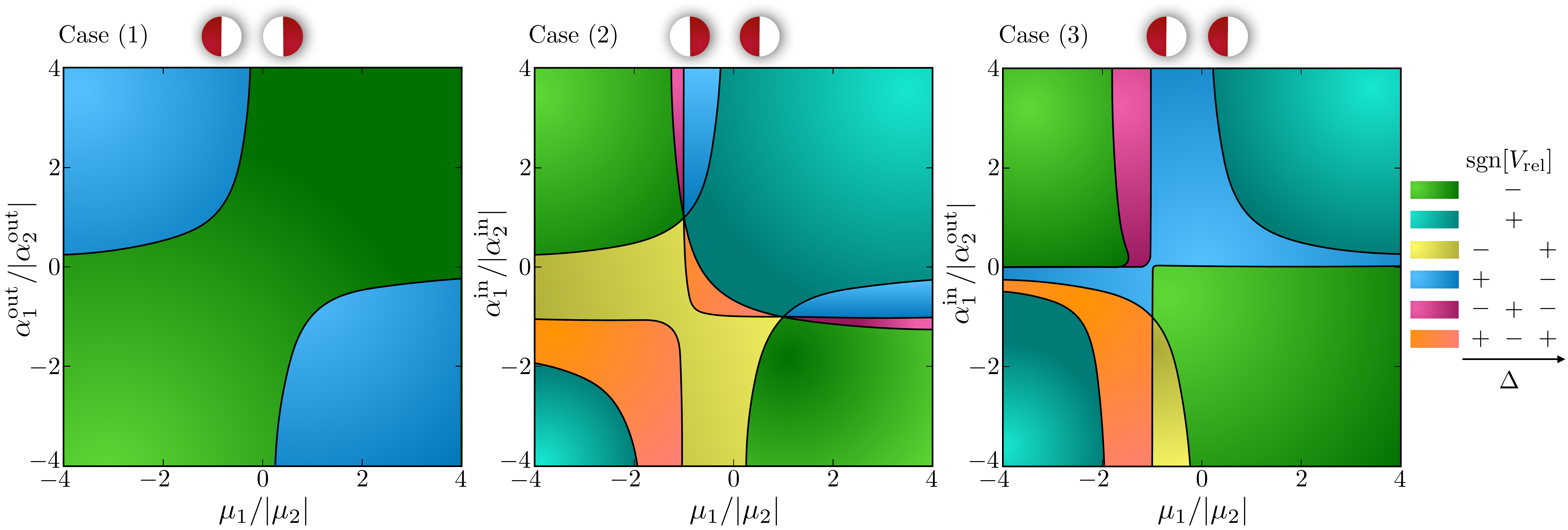}
\caption{The regime diagrams describing the relative dynamics of half-coated particles for three configurations. As shown by the schematics at the top of each panel (red and white colours represent active and inert compartments, respectively), the three configurations are: Case (1) inner compartments are inert. Case (2) outer compartments are inert. Case (3) inner compartment of sphere 1 and outer compartment of sphere 2 are inert. As shown in the right side of the figure, colours represent variations of the nature of interactions (attractive or repulsive) versus the gap size. Note that these maps must be reversed if $\alpha_2\mu_2<0$.}
\label{diagram}
\end{center}
\end{figure}
As shown in figure~\ref{v_rel_plot}, the system can have one single fixed point (stable or unstable), two fixed points (one stable, one unstable), or three fixed points (two stable, one unstable or vice versa). This means that a pair of Janus particles may exhibit a variety of behaviours, depending on their initial gap size. When the system has no fixed point, the interactions are either purely attractive in which the particles collapse and make a complex, or purely repulsive in which they separate indefinitely. A single stable fixed point indicates that the particles (regardless of their initial position) hold a nonzero gap size at steady state and subsequently move together with an identical velocity. For a single unstable fixed point, the particles form a metastable complex if their initial gap size is below a certain value, and move away if their gap size exceeds that value. The behaviour becomes more complicated once the system exhibits more than one fixed point. For the case of two fixed points, the particles reach an equilibrium state at a nonzero gap size. This state is however only linearly stable, thus, under sufficient perturbation (e.g., thermal activation) the particles either form a metastable complex (when the gap size corresponding to the stable fixed point is larger than the one of the unstable fixed point) or move away (when the gap size corresponding to the stable fixed point is smaller than the one of the unstable fixed point). When the system has three fixed points, there are two scenarios for the relative interaction. If two of these fixed points are stable, then the particles reach a steady state at a nonzero gap size. There are two stable fixed points in this case, hence this equilibrium gap size can vary between two values, and so the system can move from one state to another under the presence of a noise. In the case of two unstable and one stable fixed points, the system reaches a linearly-stable state at a nonzero gap size, and will either form a metastable complex, or separate under sufficient perturbations.

\subsection{Half-coated particles}
By using the generic expression given in \eqref{relative2}, one can simply determine the nature of the interactions for any pair of Janus particles at any gap size. Nevertheless, given the importance of half-coated particles (Janus particles with one compartment being completely inert) in the experimental realization of chemically active systems \cite{Ebbens2012,Ebbens2014,Brown2014,Ebbens2018}, it is worthwhile to further evaluate equation \eqref{relative2} for cases wherein one side of each particle is inert. We can thereby have three configurations: case (1) wherein the two inner sides are inert $\alpha_1^\text{in}=\alpha_2^\text{in}=0$, case (2) in which the inner sides are active $\alpha_1^\text{out}=\alpha_2^\text{out}=0$, and case (3) with $\alpha_1^\text{out}=\alpha_2^\text{in}=0$. For case (1) we find
\begin{align}
\label{case1}
V_\text{rel}^{(1)}=\left(\mathcal{G}^\text{self}-\mathcal{G}^\text{ref,out}\right)\left[\left(\mu_1\alpha_2^\text{out}+\mu_2\alpha_1^\text{out}\right)\varepsilon_3-\left(\mu_1\alpha_1^\text{out}+\mu_2\alpha_2^\text{out}\right)\right]/D,
\end{align}
which indicates that there can be only one fixed point in this configuration of the particles, since the variation of $\varepsilon_3$ with $\Delta$ is monotonic, as shown in figure~\ref{gs}(b). For case (2), we similarly find
\begin{align}
V_\text{rel}^{(2)}=\mathcal{G}^\text{nei,in}\left[\left(\mu_1\alpha_2^\text{in}+\mu_2\alpha_1^\text{in}\right)+\left(\mu_1\alpha_1^\text{in}+\mu_2\alpha_2^\text{in}\right){\varepsilon_2}/{\varepsilon_1}\right]/D,
\label{case2}
\end{align}
where $\varepsilon_2/\varepsilon_1$ is now a non-monotonic function with respect to $\Delta$ and so the system can have two fixed points. Finally, for case (3), we have
\begin{align}
V_\text{rel}^{(3)}=\left(\mathcal{G}^\text{self}-\mathcal{G}^\text{ref,out}\right)\left(\mu_2\alpha_1^\text{in}\varepsilon_1+\mu_1\alpha_1^\text{in}\varepsilon_2+\mu_1\alpha_2^\text{out}\varepsilon_3-\mu_2\alpha_2^\text{out}\right)/D.
\label{case3}
\end{align}
Now, using equations \eqref{case1} to \eqref{case3}, we construct the phase diagrams describing the dynamical behaviour of the particles, in the activity-mobility parameter space (see figure~\ref{diagram}). For half-coated particles, we find that the system can no longer have three fixed points.

\color{black}
\color{black}
\section{Non-uniform activity and non-uniform mobility}
\begin{figure}
\begin{center}
\includegraphics[width=0.4\columnwidth]{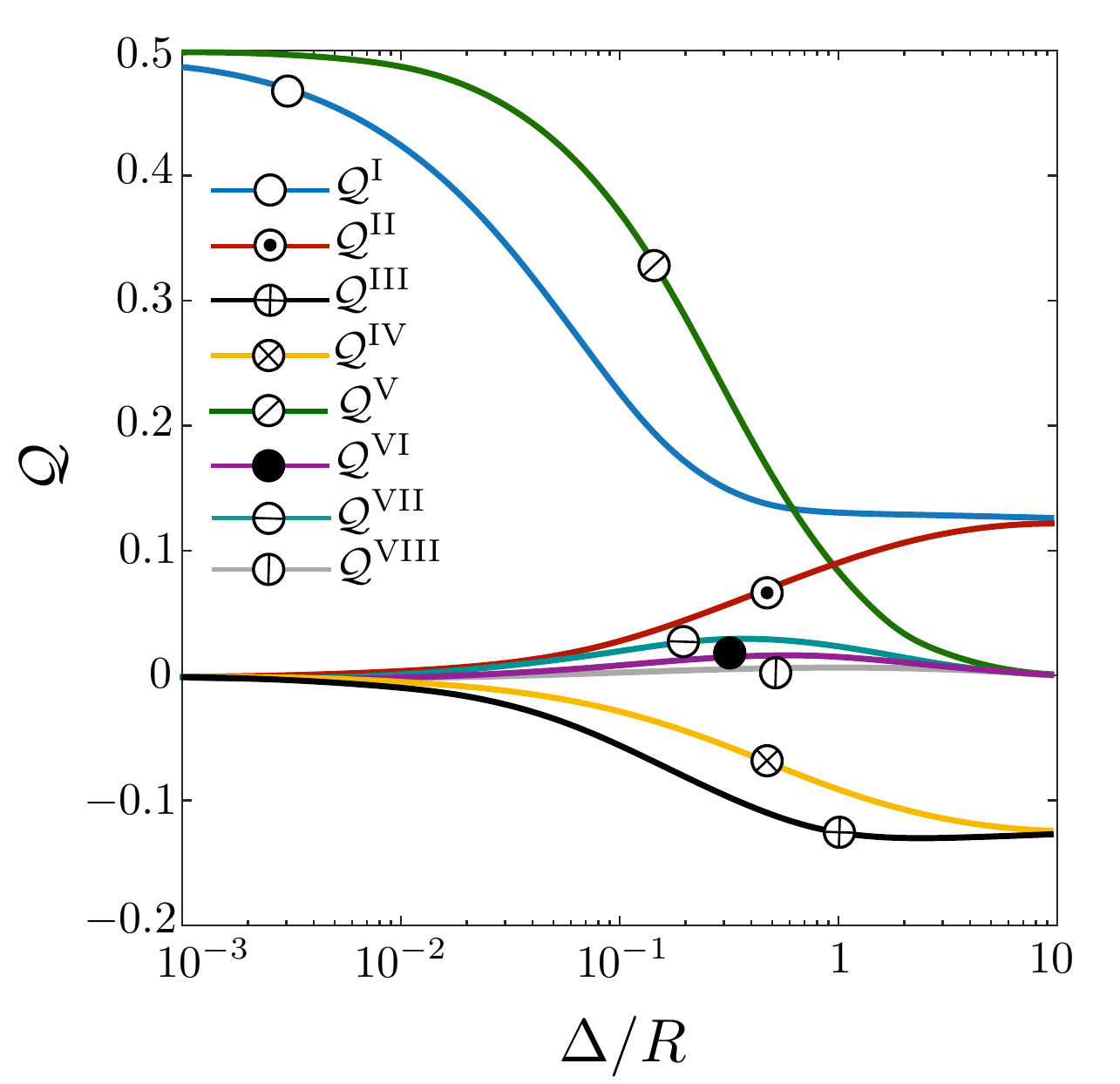}
\caption{\color{black} The variation of the geometrical $\mathcal{Q}$ functions versus the gap size $\Delta$. A linear summation of these geometrical functions can return the relative speed of two Janus particles whose compartments have different activities and mobilities, as shown in equation \eqref{relative-gen}.}
\label{g-gen}
\end{center}
\end{figure}

We now look at the general case in which the two compartments of each particle can have different values of activities and mobilities. In this case, the neighbour-induced and the neighbour-reflected motions are more entangled, thus the dimensionless geometrical functions should be defined more generally as
\begin{align}
\frac{D}{|\hat{\bm{F}}_1|}\left<\mu_i\hat{\sigma}_i\bm{\nabla}^i_\parallel C_i\right>_{\mathcal{S}_i}&=\bm{e}\mu_i^\text{in}\alpha_i^\text{in}\mathcal{Q}^\text{I}_i+\bm{e}\mu_i^\text{out}\alpha_i^\text{in}\mathcal{Q}^\text{II}_i+\bm{e}\mu_i^\text{in}\alpha_i^\text{out}\mathcal{Q}^\text{III}_i+\bm{e}\mu_i^\text{out}\alpha_i^\text{out}\mathcal{Q}^\text{IV}_i,\\
\frac{D}{|\hat{\bm{F}}_1|}\left<\mu_i\hat{\sigma}_i\bm{\nabla}^i_\parallel C_j\right>_{\mathcal{S}_i}&=\bm{e}\mu_i^\text{in}\alpha_j^\text{in}\mathcal{Q}^\text{V}_i+\bm{e}\mu_i^\text{out}\alpha_j^\text{in}\mathcal{Q}^\text{VI}_i+\bm{e}\mu_i^\text{in}\alpha_j^\text{out}\mathcal{Q}^\text{VII}_i+\bm{e}\mu_i^\text{out}\alpha_j^\text{out}\mathcal{Q}^\text{VIII}_i,
\end{align}
Again, under a mirror-symmetric transformation we have  $\left(\mathcal{Q}_i\right)^*=\mathcal{Q}_j\equiv\mathcal{Q}$, thus the relative velocity this time is found
\begin{align}
\label{relative-gen}
V_\text{rel}=&\mathcal{Q}^\text{I}\left(\mu_1^\text{in}\alpha_1^\text{in}+\mu_2^\text{in}\alpha_2^\text{in}\right)/D+\mathcal{Q}^\text{II}\left(\mu_1^\text{out}\alpha_1^\text{in}+\mu_2^\text{out}\alpha_2^\text{in}\right)/D+\mathcal{Q}^\text{III}\left(\mu_1^\text{in}\alpha_1^\text{out}+\mu_2^\text{in}\alpha_2^\text{out}\right)/D+\nonumber\\
&\mathcal{Q}^\text{IV}\left(\mu_1^\text{out}\alpha_1^\text{out}+\mu_2^\text{out}\alpha_2^\text{out}\right)/D+\mathcal{Q}^\text{V}\left(\mu_1^\text{in}\alpha_2^\text{in}+\mu_2^\text{in}\alpha_1^\text{in}\right)/D+\mathcal{Q}^\text{VI}\left(\mu_1^\text{out}\alpha_2^\text{in}+\mu_2^\text{out}\alpha_1^\text{in}\right)/D+\nonumber\\
&\mathcal{Q}^\text{VII}\left(\mu_1^\text{in}\alpha_2^\text{out}+\mu_2^\text{in}\alpha_1^\text{out}\right)/D
+\mathcal{Q}^\text{VIII}\left(\mu_1^\text{out}\alpha_2^\text{out}+\mu_2^\text{out}\alpha_1^\text{out}\right)/D.
\end{align}
Comparing this solution to the one of uniform mobility given in equation \eqref{relative2}, it is clear that $\mathcal{Q}^\text{I}+\mathcal{Q}^\text{II}=\mathcal{G}^\text{ref,in}+\mathcal{G}^\text{self}$, $\mathcal{Q}^\text{III}+\mathcal{Q}^\text{IV}=\mathcal{G}^\text{ref,out}-\mathcal{G}^\text{self}$, $\mathcal{Q}^\text{V}+\mathcal{Q}^\text{VI}=\mathcal{G}^\text{nei,in}$, and $\mathcal{Q}^\text{VII}+\mathcal{Q}^\text{VIII}=\mathcal{G}^\text{nei,out}$. Thus, under this decomposition, the self-propulsion term is distributed between $\mathcal{Q}^\text{I}$ to $\mathcal{Q}^\text{IV}$, as one can also see in figure~\ref{g-gen} since they are the only geometrical functions that do not decay to zero as the gap size increases. One can alternatively separate the self-propulsion from the interaction-induced terms as here we have \cite{golestanian2007}
\begin{align}
\frac{D}{|\hat{\bm{F}}_1|}\left<\mu_i\hat{\sigma}_i\bm{\nabla}^i_\parallel C^\text{far}_i\right>_{\mathcal{S}_i}&=\bm{e}\mathcal{G}^\text{self}_i\left(\frac{\mu_i^\text{in}+\mu_i^\text{out}}{2}\right)\left(\alpha_i^\text{in}-\alpha_i^\text{out}\right).
\end{align}
Similar to the case of uniform mobilities, here as well as in the lubrication regime, the effect of the outer compartments to the relative motion becomes irrelevant. As shown in figure~\ref{g-gen}, when $\Delta\rightarrow0$, only $\mathcal{Q}^\text{I}$ and $\mathcal{Q}^\text{IV}$ have nonzero values indicating that even the effect of {cross inner-outer} terms such as $\mathcal{Q}^\text{II}$ and $\mathcal{Q}^\text{III}$ vanish away in this limit. 

{\color{black}{We perform a thorough parameter scan over the activity-mobility parameter space to identify the emergence of fixed points in the system. Surprisingly, we find that allowing the mobilities to be nonuniform across the surface of the particles does not increase the number of fixed points in the dynamical system. This may be due to the fact that unlike the chemical activities that alter the chemical field directly, any discontinuity in the mobilities can only directly affect the hydrodynamic field which has proven to be less important in terms of determining the fixed points in the dynamical system \cite{nasouri2020}.}} Thus, we can then conclude that a pair of Janus particles can only have up to three fixed points in their relative motion.

\color{black}
\section{Conclusion}
In this study, we discussed the {\color{black}{axisymmetric}} pair interaction of two Janus particles, and derived a generic solution for their relative motion. {\color{black}{This solution, which is in terms of a linear summation of geometrical functions, illustrates the contribution of each compartments of the particles to the relative motion}}. Since in far-field based many-body solvers the near-field effects are often taken into the account through pair interactions, the generic solution presented here can in particular provide an efficient and accurate way to introduce near-field effects when modelling phoretic suspensions. We also use this solution to show that the dynamical system describing the relative motion can have up to three fixed points, indicating that the system can exhibit vastly different behaviours depending on the initial gap size and the chemical properties of the particles. 

Because of its simplicity and generality, our approach can be simply extended to study the interaction of Janus particles in which the coating coverage of the two compartments are not identically equal. For these systems, if the front-back geometrical symmetry is broken, one needs to keep more geometrical functions to construct the generic solution. The geometrical asymmetry in these cases then may induce more fixed points in the system. Similarly, if the particles have more complicated coating patterns, one may also expect a higher number of fixed points to emerge. {\color{black}{The calculation can also be extended to cases where the particles have slender axisymmetric shapes \cite{ibrahim2018} and cases where more experimentally relevant details of the chemical reaction are taken into consideration \cite{ibrahim2017}. For instance, we can similarly introduce these geometrical functions for interactions of two spheroidal particles. Given that the exact motion of a single spheroidal squirmer with a catalytic surface has been recently discussed by \cite{pohnl2020}, one can use that solution to construct a reflection-based approach for capturing the far-field behaviour of two spheroidal particles. However, studying the full behaviour of the system still requires an exact evaluation of the geometrical function for which computational approaches (such as the boundary element method \cite{uspal2019}) should be employed.}}

{\color{black}{Furthermore, we should note that to evaluate the stability of the reported bound states, one should look into non-axisymmetric interactions, since our current axisymmetric approach does not take into account rotation and lateral translation which could trigger an escape under the presence of a noise. In that case, other than the gap size, the geometrical functions also depend on the orientation of the particles. Thus, at a given gap size, these functions should be evaluated for all the possible orientations. Given the cumbersomeness of these calculations for non-axisymmetric cases \cite{sharifi2016}, it may be useful to limit the exact treatment of the geometrical function to only small gap sizes and use the far-field approach when the particles are far from one another.}}

\color{black}
We also note that advective effects of the solute particles, which are neglected here, can induce similar stable and unstable fixed points in the dynamical system \cite{lippera2020}. Thus, one may adapt the presented approach to identify the possible scenarios for the relative motion when the P\'eclet number is not identically zero.


%


\end{document}